\documentclass[conference]{IEEEtran}
\IEEEoverridecommandlockouts
\usepackage{cite}
\usepackage{amsmath,amssymb,amsfonts}
\usepackage{algorithmic}
\usepackage{graphicx}
\usepackage{textcomp}
\usepackage{xcolor}
\usepackage{caption}
\usepackage{subcaption}
\usepackage{placeins}
\usepackage{array}
\usepackage{multirow}
\usepackage{makecell}
\def\BibTeX{{\rm B\kern-.05em{\sc i\kern-.025em b}\kern-.08em
    T\kern-.1667em\lower.7ex\hbox{E}\kern-.125emX}}
\begin{document}

\title{Cloud-Connected Wireless Holter Monitor with Neural Networks Based ECG Analysis for Remote Health Monitoring}

\author{\IEEEauthorblockN{1\textsuperscript{st} Azlaan Ranjha}
\IEEEauthorblockA{\textit{Department of Electrical Engineering} \\
\textit{NUST College of EME}\\
Rawalpindi, Pakistan \\
azranjha.ee43ceme@student.nust.edu.pk}
\and
\IEEEauthorblockN{2\textsuperscript{nd} Laiba Jabbar}
\IEEEauthorblockA{\textit{Department of Electrical Engineering} \\
\textit{NUST College of EME}\\
Rawalpindi, Pakistan \\
ljabbar.ee43ceme@student.nust.edu.pk}
\and
\IEEEauthorblockN{3\textsuperscript{rd} Osaid Ahmad}
\IEEEauthorblockA{\textit{Department of Electrical Engineering} \\
\textit{NUST College of EME}\\
Rawalpindi, Pakistan \\
oahmad.ee42ceme@student.nust.edu.pk}
}
\maketitle

\begin{abstract}
This study describes the creation of a wireless, transportable Holter monitor to improve the accuracy of cardiac disease diagnosis. The main goal of this study is to develop a low-cost cardiac screening system suited explicitly for underprivileged areas, addressing the rising rates of cardiovascular death. The suggested system includes a wireless Electrocardiogram (ECG) module for real-time cardiac signal gathering using attached electrodes, with data transfer made possible by WiFi to a cloud server for archival and analysis. The system uses a neural network model for automated ECG classification, concentrating on the identification of cardiac anomalies. The diagnostic performance of cardiologist-level ECG analysis is surpassed by our upgraded deep neural network architecture, which underwent thorough evaluation and showed a stunning accuracy rate of more than 88\%. A quick, accurate, and reasonably priced option for cardiac screening is provided by this ground-breaking technology, which smoothly merges wireless data transfer with AI-assisted diagnostics. In addition to providing a thorough overview of the development process, this paper also highlights methods used to improve model accuracy, such as data preparation, class imbalance correction using oversampling, and model fine-tuning. The work shows the viability of a comprehensive remote cardiac screening system powered by AI and maximising the use of wearable and cloud computing resources. Such cutting-edge remote health monitoring technologies have great promise for improved health outcomes and early identification, especially in resource-constrained countries.

\end{abstract}

\section{Introduction}
Cardiovascular disorders, which include ailments like myocardial infarction, cardiac arrest, and arrhythmia, continue to be the largest cause of death worldwide. To effectively identify and manage these illnesses, electrocardiogram (ECG) analysis has emerged as a key technique. Medical professionals can identify anomalies by using an ECG, which examines the complex electrical signals driving heart rhythm.

Traditional Holter monitoring requires wearing ECG equipment continuously for several days, followed by post-recording analysis. Real-time wireless ECG data transmission has become possible because of technological advancements, which make remote monitoring easier. Automating ECG analysis using deep neural networks enables quick screening and diagnosis of heart abnormalities. Through the use of wireless signal transmission and deep learning, this project aims to develop an intelligent wireless Holter monitoring system that will provide quick, reliable, and affordable cardiac screening.

The suggested methodology calls for the wireless real-time transmission of ECG signals from wearable sensors to cloud-based computers. The wireless data transmission is to be done by the use of WiFi modules such as the ESP32. The selection of an ESP32 as a WiFi module was done as they are low-cost and extremely versatile, therefore they are recommended for such projects. A neural network model then performs automated classification to separate between healthy and unhealthy cardiac beats. With the use of AI-powered remote cardiac monitoring, this programme aims to reduce the rising worldwide burden of cardiovascular diseases by providing the possibility of early detection and preventive intervention.

The integrated wireless ECG system shown in this work, along with a deep learning algorithm, is intended to improve and scale heart health screening. This paper explains in great detail the system design, implementation, and real-world evaluation.

\section{Problem Statement}\label{SCM}

With over 17 million fatalities reported in 2019, cardiovascular diseases (CVDs) have emerged as a global health issue and are now the leading cause of death worldwide\cite{b10}\cite{b11}. Notably, low and middle-income countries endure a disproportionate amount of the burden, with CVDs serving as these areas' main cause of mortality\cite{b10}. For instance, in Pakistan alone in 2020, 240,720 people died as a result of CVDs, accounting for more than 16\% of all fatalities that year\cite{b12}. Myocardial infarction and strokes are to blame for the majority of these CVD-related deaths\cite{b10}\cite{b13}. The burden of CVD-related mortality can be reduced with quick and precise diagnosis combined with prevention actions\cite{b14}. In this context, electrocardiogram (ECG) monitoring is crucial since it enables medical professionals to spot abnormal heart rhythms linked to CVDs\cite{b15}\cite{b16}. However, classic Holter ECG devices have some serious disadvantages, such as obtrusive electrical connections and skin discomfort from prolonged use\cite{b17}\cite{b18}. Furthermore, real-time wireless data transmission is not possible with conventional Holters, which might cause delays in diagnosis and treatment\cite{b19}. Recent developments have allowed us to provide a small, wireless ECG Holter monitor that makes use of cloud-based computing and artificial intelligence, offering potential answers\cite{b20}\cite{b21}\cite{b22}\cite{b23}. The suggested system is proposed to use wireless wearable sensors to record real-time electrocardiogram (ECG) signals and send the information to cloud servers for distant data archiving and analysis. Based on the wireless data acquired, a powerful deep neural network model automatically distinguishes between normal and pathological heartbeats. This ground-breaking method can potentially` democratise affordable and convenient cardiac screening by removing geographical restrictions\cite{b21}. Furthermore, AI-powered diagnosis makes it easier to quickly identify potentially fatal cardiac anomalies\cite{b20}\cite{b23}. This study aims to develop and verify a cloud-based deep learning analytics-based integrated wireless ECG monitoring system\cite{b21}. If it works, this sophisticated remote cardiac screening method could be a game-changer in the early detection and prevention of CVD\cite{b21}. This strategy, especially in areas with limited resources, shows promise in reducing the rising burden of CVD-related mortality by eliminating barriers that prevent prompt life-saving therapies\cite{b21}.

\section{Literature Review}
An in-depth knowledge of current research is essential to advancing solutions and addressing the rising global burden of cardiovascular diseases (CVDs) in the fields of remote health monitoring and intelligent medical systems. This literature review examines current research on wireless Electrocardiogram (ECG) devices and deep learning for automated cardiac arrhythmia detection, highlighting significant improvements and unresolved problems that warrant additional study. This study seeks to provide context and motivate the AI-powered wireless Holter monitor presented in this research by carefully reviewing pertinent papers published in prestigious journals and conference proceedings. We will evaluate current methods for cloud-based analysis, remote ECG capture, and machine-learning models for accurate ECG classification. To advance the state-of-the-art in accessible and accurate cardiac screening through intelligent remote health monitoring systems, it is important to comprehensively evaluate the ideas that are most relevant to our research while also identifying interesting directions. The design choices we make for our methodologies and the contextualization of our contributions will be influenced by this foundation.

\subsection{Issues with traditional Holter monitors}\label{AA}
Numerous studies have repeatedly concluded that typical Holter monitors have significant shortcomings when it comes to the identification of cardiovascular disease. These restrictions show up as a scant diagnostic yield that typically ranges between 15\% to 40\% \cite{b1}\cite{b9}, with around 23\% of patients requiring additional Holter testing. This inefficiency grows more pressing in the United States, where a single patient testing session costs a significant \$23,000. Furthermore, these conventional instruments' lengthy diagnostic turnaround times worsen the problem. As a result, a plethora of substitute products have hit the market, however, they tend to concentrate primarily on ECG data and frequently ignore the important impact that patients' background demographics play, which has been noted in the literature \cite{b9}. Medical professionals who may not have the data analysis skills displayed by Machine Learning (ML) models are commonly trusted with handling these demographic intricacies. Furthermore, given that doctors can only analyse the data after around 24-48 hours have passed, standard Holter monitors struggle with ongoing issues such as short battery life. Additionally, there are the issues of skin discomfort brought on by surface electrode leads, and delays in diagnosis \cite{b3}. Additionally, there is cause for worry because about a quarter (25\%) of patients have trouble starting their devices while they are having symptomatic episodes\cite{b1}.

\subsection{ECG Performance and WiFi Data Transmission}
The use of the ESP32 for wireless data transmission has been determined to be the best option due to its affordability and effective data transfer capabilities, assuring the preservation of data quality and reducing the danger of data loss. Researchers have looked into using a Raspberry Pi computer to receive transmitted data in some cases, and this has shown the capacity to sustain high-quality data transmission \cite{b2}.

Devices like SmartCardia \cite{b1} and Zio \cite{b9}, each offering particular benefits, are readily available on the market. Through the use of a specialised adhesive patch, SmartCardia tackles the issue of skin sensitivity. Zio, on the other hand, has bragged about how accurately it gathers data. It's crucial to recognise the research's limitations, though. Only 40 candidates were tested a rather small sample size, which is insufficient for a thorough evaluation \cite{b1}. In addition, SmartCardia makes use of a Bluetooth module, which can present its own set of difficulties and risks in the context of a complex health-focused solution \cite{b1}.

\subsection{Low Power Consumption}
Power consumption is a key issue when it comes to wireless or battery-operated gadgets. Implementing a sleep mode during periods of inactivity and optimising hardware to maximise power efficiency are two strategies that may be used to address this problem \cite{b2}.

\subsection{Alternate Methods of Cardiovascular Diagnosis}
Different scientific disciplines have investigated other ways of evaluating cardiac abnormalities in addition to electrocardiogram (ECG) data. Pharmacological treatments for cardiac diseases can be just as successful as surgical ones, according to research. Pharmacological techniques have occasionally been effective in lowering the risk of abrupt arrhythmia. Additionally, certain genetic variants linked to a higher risk of cardiac arrest have been found\cite{b5}.

Diverse modelling approaches have been employed by researchers to enhance understanding within the medical community regarding the relationship between cardiac arrhythmia and heart failure. These encompass clinical cases (human patient data), animal testing (lab experiments involving animals), pluripotent stem cell testing, and computational models. These methodologies offer valuable insights, aiding medical practitioners in comprehending heart diseases and the influence of various demographic factors on heart-related conditions. Specifically, in the mentioned research, calcium dyes were utilized to induce heart contractions. Subsequently, these contraction patterns were employed to train machine learning models for patient diagnosis \cite{b4}.

Convolutional Neural Networks (CNNs), a type of machine learning model, have emerged as the optimal choice for ECG data analysis due to their superior model accuracy. In a comparative analysis, CNNs outperformed models such as BP Neural Networks and Random Forest Classifiers, achieving an accuracy rate of approximately 99.7\% based on key parameters including accuracy, sensitivity, specificity, and positive prediction rate. The models effectively classified the results into multiple categories, demonstrating robustness and reliable performance\cite{b7}. However, it is crucial to note that the model’s reliance solely on ECG data limits its diagnostic accuracy. Furthermore, the results were entirely software-based, excluding any potential variations introduced by hardware integration. Despite these limitations, the model’s architecture shows significant potential and can be adapted to develop more sophisticated and personalized models for improved task performance.

\subsection{Comparison with Human Cardiologist}\label{SCM}
The core of our research is the possibility for our model to outperform human medical professionals in terms of diagnostic precision. This capability serves as the foundation for the benefits and justification of creating such a model, which aims to support cardiologists and other medical professionals in making accurate judgements.

According to studies, the average healthcare professional's accuracy is around 54\%, but with the right training, that figure may rise to about 67\%\cite{b8}. However, this figure is very low, especially in a crucial sector like cardiology where there is little room for error because of the high risk of death. According to additional data, trained cardiologists, residents, and medical students diagnose ECGs with an accuracy rate of about 42\%, 68.5\%, and 74\%, respectively\cite{b8}.

While additional training could potentially enhance a cardiologist’s accuracy in ECG reading, this approach is financially burdensome and time-intensive\cite{b8}. Furthermore, there is no assurance of universal improvement. Consequently, there has been a surge in research focusing on machine-learning models for cardiac ailments, given their superior accuracy, precision, and efficiency\cite{b8}.

\section{Methodology}
\subsection{Machine Learning Model}
\subsubsection{Dataset Collection}
Our research project's first stage involved the thorough collection of reliable datasets. Numerous cardiologists were consulted in-depth and interviewed before the process of data gathering began. These consultations had the goal of clarifying the key characteristics required for correctly identifying cardiac arrhythmia and other heart disorders in patients. It is crucial to stress that real patient data was not used until the appropriate parties gave their consent, due to ethical reasons. Instead, our dataset was meticulously sourced from reputable and established online repositories and sources with a proven track record of data quality and reliability. Subsequently, a comprehensive amalgamation of various datasets was executed to construct a more expansive and tailored dataset aligned with our testing requirements. This culminated in the creation of the final merged dataset, comprising categorical data derived from patients' medical histories and background demographics \cite{b24}, in conjunction with ECG data obtained from patients \cite{b25}.

\begin{figure}[htbp]
  \centering
  \includegraphics[width=0.8\linewidth]{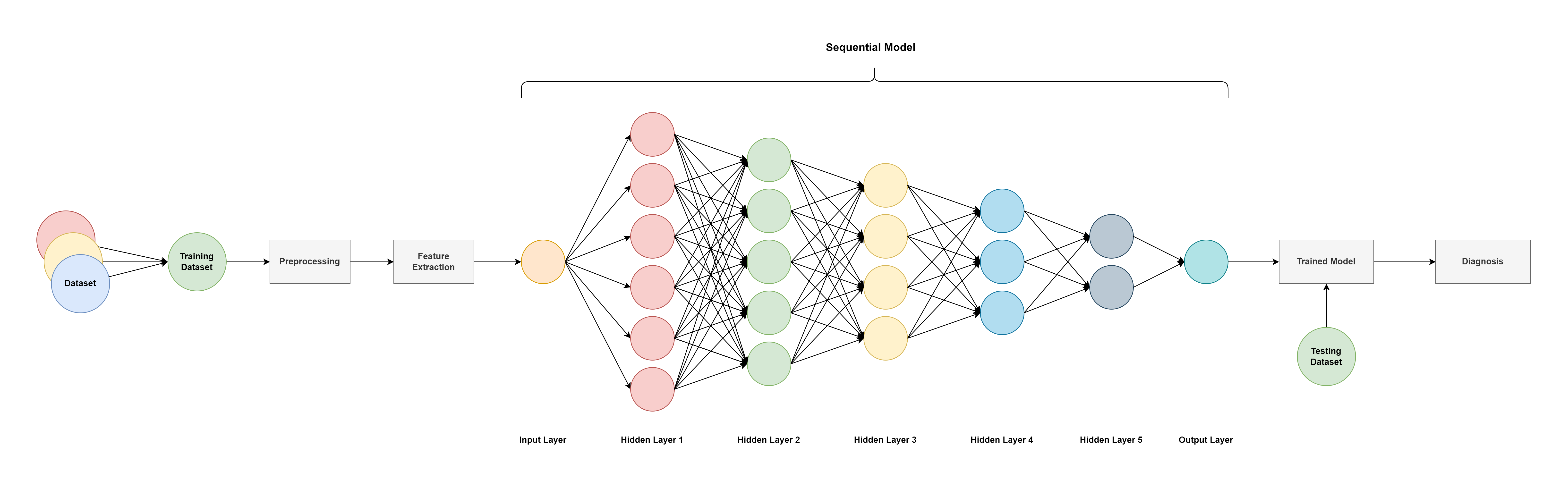}
  \caption{Machine Learning Pipeline}
  \label{fig:ml_pipeline}
    \captionsetup{font=small} % Set the font size to small for the subcaption
  \subcaption{Many datasets are merged together to form a single dataset. After data preprocessing and feature extraction, this dataset is sent into a sequential model of 7 dense neural network layers. The final trained model with about 89\% accuracy is then used on unseen test data to give a diagnosis of whether the patient is normal or has an abnormal heart condition.}
\end{figure}

\subsubsection{Data Preprocessing}
The final amalgamated dataset underwent a rigorous categorization process, resulting in two distinct classes of patient labels: "Normal Heart Condition" and "Abnormal Heart Condition." It is noteworthy that the ECG data was initially organized into separate folders denoting normal and abnormal data, yet lacked corresponding labels. To address this, a deliberate integration strategy was employed, whereby normal ECG data was concatenated with the records of patients classified as "normal," and conversely, abnormal ECG data was linked with the corresponding "abnormal" patient records. Following this integration, the dataset underwent a systematic shuffling procedure, accompanied by a resetting of data indices to ensure randomness and uniformity.

\begin{figure}[htbp]
  \centering
  \captionsetup{font=small} 
  \begin{subfigure}{0.45\linewidth}
    \includegraphics[width=\linewidth]{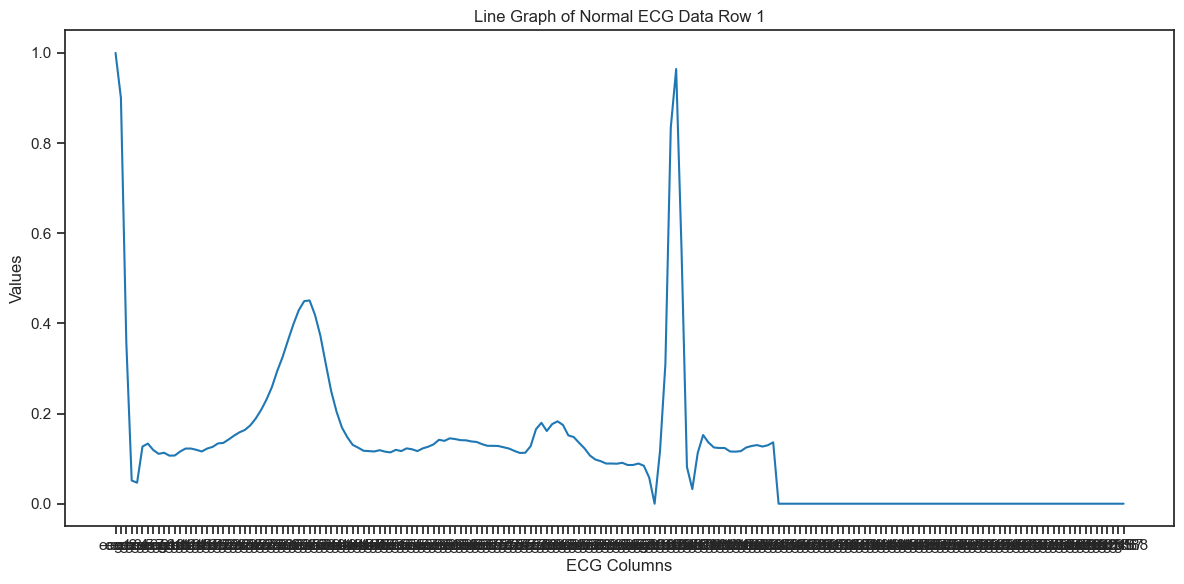}
    \caption{Normal Patients ECG}
    \label{fig:normal_ecg}
  \end{subfigure}
  \hfill
  \begin{subfigure}{0.45\linewidth}
    \includegraphics[width=\linewidth]{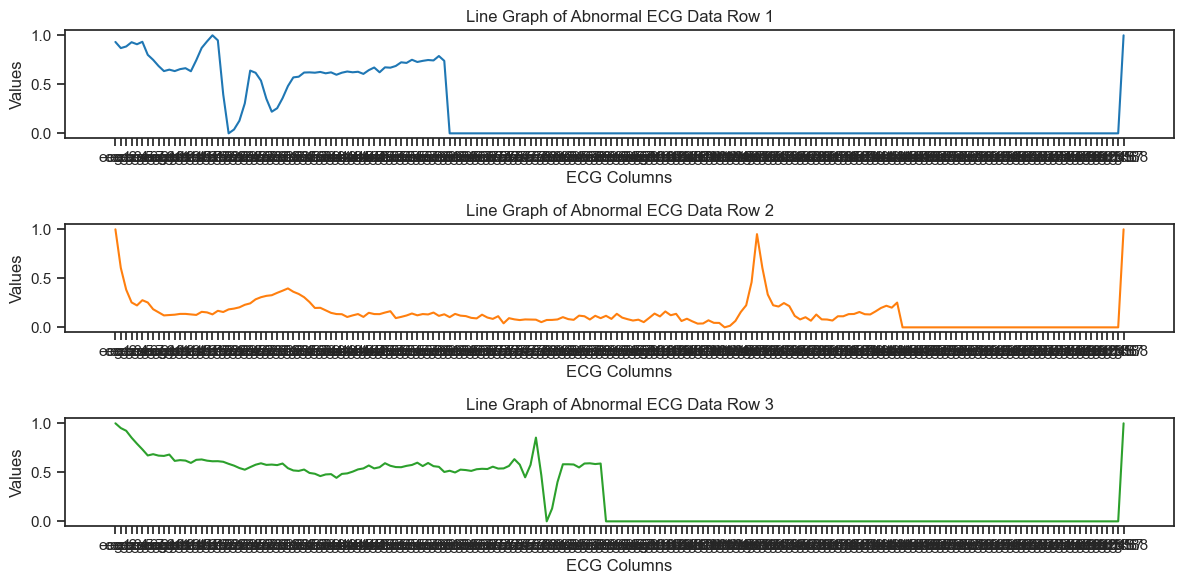}
    \caption{Abnormal Patients ECG}
    \label{fig:abnormal_ecg}
  \end{subfigure}
  
  \vspace{\baselineskip} % Add some vertical space between rows
  
  \begin{subfigure}{0.45\linewidth}
    \includegraphics[width=\linewidth]{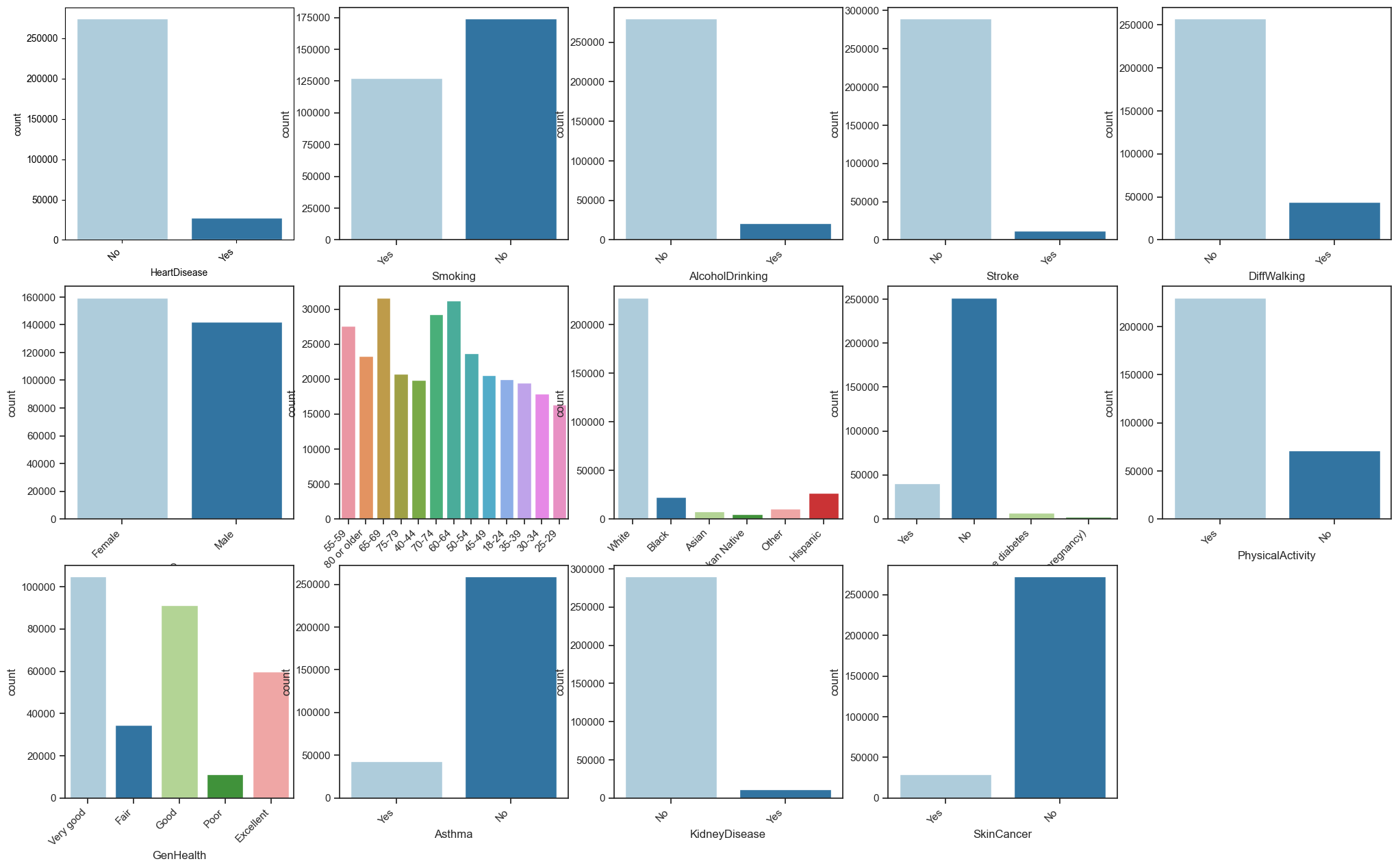}
    \caption{Categorical Data Features Distribution}
    \label{fig:categorical_distribution}
  \end{subfigure}
  \hfill
  \begin{subfigure}{0.45\linewidth}
    \includegraphics[width=\linewidth]{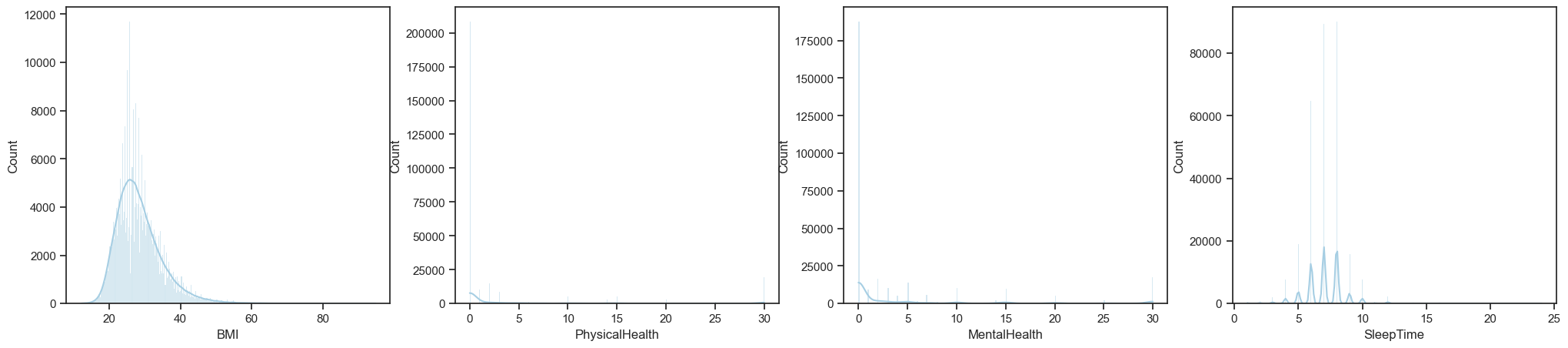}
    \caption{Numeric Data Features Distribution}
    \label{fig:numeric_distribution}
  \end{subfigure}
  \caption{Dataset Figures}
  \label{fig:combined}
\end{figure}

Both before and following the integration process, meticulous data quality control measures were implemented. These included regular inspections and the removal of any instances of null values, duplicate entries, or outliers. Furthermore, to enhance the fidelity of the ECG data, Butterworth Lowpass filters were applied to mitigate noise artefacts. These filters, widely adopted in signal processing, are favoured for their characteristics, which encompass a flat pass-band response, gradual roll-off, ease of design, and stability in steady-state operation. It is imperative to acknowledge that the choice of filter type hinges on specific application requirements, with alternative options such as Chebyshev, elliptic, and Finite Impulse Response (FIR) filters offering varying trade-offs to cater to distinct needs\cite{b26}\cite{b27}\cite{b28}.
\begin{figure}[htbp]
  \centering
  \includegraphics[width=200px, height=100px]{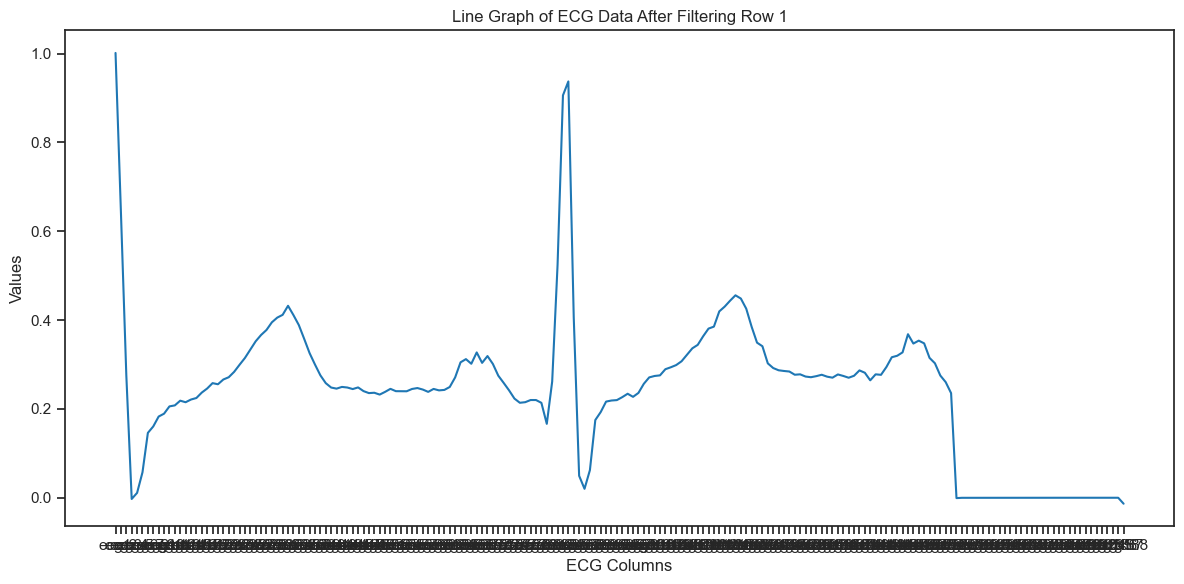}
  \caption{Filtered Normal ECG Signal}
  \label{fig}
\end{figure}

\subsubsection{Feature Engineering}
Following the meticulous completion of the Data Preprocessing phase within our model training pipeline, we transitioned to the subsequent step, namely, feature engineering. Leveraging the rich ECG data at our disposal, we embarked on a journey to expand and diversify our dataset, thereby enhancing our model's training regimen and its ability to gain deeper insights into the patients' conditions.

For each column of the ECG data, a comprehensive array of features was diligently extracted and subsequently incorporated as new columns appended to the dataset. Each row in the dataset was thereby endowed with a unique set of features, encompassing parameters such as 'HR Variability,' 'QRS Duration Mean,' 'QRS Duration STD,' 'Spectral Sum,' 'Spectral Max,' 'Spectral Max Frequency,' 'Spectral Mean,' 'Spectral STD,' 'Skewness,' 'Kurtosis,' and many more. This enrichment process served to provide the model with a more comprehensive and nuanced representation of the underlying ECG data\cite{b29}\cite{b30}.

Subsequently, the datasets underwent a normalization procedure, and steps were taken to rectify any imbalances present within the label classes. Following a final round of data filtration, the dataset was judiciously partitioned into distinct training and test sets, primed for deployment in our training model\cite{b31}\cite{b32}.

\subsubsection{Model Structure and Training}
An extremely nuanced and complex sequential training model was designed using tools such as TensorFlow and Keras.\cite{b33} The model has 7 Dense layers, along with the output layer. After each layer, normalization and dropout methods are used\cite{b34}\cite{b35}\cite{b36}. Alongside these, other elements such as  Learning Rate Scheduler, Callbacks and Early Stopping are also used\cite{b37}\cite{b38}. All this has been done to increase model accuracy and reduce over-fitting and model losses. As our classification labels were binary encoded we used binary cross-entropy loss function\cite{b39}\cite{b40}\cite{b41}\cite{b42} and many other metrics such as accuracy and AUC score etc\cite{b43}. were used to guide the model towards our desired functionality. 

\subsubsection{Model Results}
The dataset was subjected to a meticulous partitioning process, with 70\% of the data allocated for training purposes and the remaining 30\% reserved for rigorous testing. To maintain a balanced representation of classes within the dataset, the 'stratify' parameter was thoughtfully utilized in association with the classification labels. Additionally, within the confines of the training data subset, a prudent allocation of approximately 10\% was set aside for validation purposes.

Upon completion of the arduous model training process, the training data yielded a noteworthy accuracy rate of approximately 88\%, affirming the model's robust performance.

\begin{table}[htbp]
  \centering
  \caption{Train Data Results}
  \begin{tabular}{lcccc}
    \hline
    Label & Precision & Recall & F1-Score & Support \\
    \hline
    0 & 0.91 & 0.84 & 0.88 & 6593 \\
    1 & 0.86 & 0.92 & 0.89 & 6636 \\
    \hline
    Accuracy & & & 0.88 & 13229 \\
    Macro Average & 0.89 & 0.88 & 0.88 & 13229 \\
    Weighted Average & 0.89 & 0.88 & 0.88 & 13229 \\
    \hline
  \end{tabular}
  \\
  \subcaption{Train Data Classification Report}
  \label{tab:train_classification_report}

  \centering
  \begin{tabular}{cccc|c}
    \multicolumn{2}{c}{} & \multicolumn{2}{c}{Predicted Labels} & \\
    \cline{3-4}
    \multicolumn{1}{c}{} & \multicolumn{1}{c}{} & Not Heart Disease & Heart Disease & \multicolumn{1}{c}{Total} \\
    \cline{2-5}
    & \makecell{Not Heart \\ Disease} & 6117 & 519 & 6636 \\
    & \makecell{Heart \\ Disease} & 1031 & 5562 & 6593 \\
    \cline{2-5}
    & Total & 7148 & 6081 & 13229 \\
  \end{tabular}
  \subcaption{Train Data Confusion Matrix}
  \label{tab:train_confusion_matrix}
\end{table}

Furthermore, it is worth highlighting that the accuracy achieved with the test data also closely approximated 88\%. This noteworthy consistency in accuracy is particularly striking given the substantial disparity in sample sizes between the training and test sets. It is a testament to the model's commendable ability to maintain its accuracy, showcasing its proficiency in effectively generalizing to previously unseen data. This robust performance underscores the model's suitability for practical applications in real-world contexts.

A crucial attribute of the model is its resilience against common issues such as under-fitting and over-fitting, further solidifying its effectiveness. This quality enhances its reliability and suitability for a wide range of practical scenarios.

\begin{table}[htbp]
  \centering
   \caption{Test Data Results}
  \begin{tabular}{lcccc}
    \hline
    Label & Precision & Recall & F1-Score & Support \\
    \hline
    0 & 0.91 & 0.84 & 0.88 & 3151 \\
    1 & 0.85 & 0.92 & 0.88 & 3150 \\
    \hline
    Accuracy & & & 0.88 & 6301 \\
    Macro Average & 0.88 & 0.88 & 0.88 & 6301 \\
    Weighted Average & 0.88 & 0.88 & 0.88 & 6301 \\
    \hline
  \end{tabular}
  \subcaption{Test Data Classification Report}
  \label{tab:test_classification_report}

  \centering
    \begin{tabular}{cccc|c}
      \multicolumn{2}{c}{} & \multicolumn{2}{c}{Predicted Labels} & \\
      \cline{3-4}
      \multicolumn{1}{c}{} & \multicolumn{1}{c}{} & Not Heart Disease & Heart Disease & \multicolumn{1}{c}{Total} \\
      \cline{2-5}
      & \makecell{Not Heart \\ Disease} & 2889 & 261 & 3150 \\
      & \makecell{Heart \\ Disease} & 492 & 2659 & 3151 \\
      \cline{2-5}
      & Total & 3381 & 2920 & 6301 \\
    \cline{2-5}
    \end{tabular}
    \subcaption{Test Data Confusion Matrix}
  \label{tab:test_confusion_matrix}
\end{table}

\FloatBarrier

Lastly, the model underwent rigorous evaluation utilizing our validation dataset. This pivotal phase serves as an invaluable introspection into the real-world capabilities and performance of our established machine-learning model.

\begin{table}[htbp]
  \centering
     \caption{Validation Data Results}
  \begin{tabular}{lcccc}
    \hline
    Label & Precision & Recall & F1-Score & Support \\
    \hline
    0 & 0.92 & 0.84 & 0.88 & 756 \\
    1 & 0.84 & 0.92 & 0.88 & 714 \\
    \hline
    Accuracy & & & 0.88 & 1470 \\
    Macro Average & 0.88 & 0.88 & 0.88 & 1470 \\
    Weighted Average & 0.88 & 0.88 & 0.88 & 1470 \\
    \hline
  \end{tabular}
   \subcaption{Validation Data Classification Report}
  \label{tab:validation_classification_report}

  \centering
    \begin{tabular}{cccc|c}
      \multicolumn{2}{c}{} & \multicolumn{2}{c}{Predicted Labels} & \\
      \cline{3-4}
      \multicolumn{1}{c}{} & \multicolumn{1}{c}{} & Not Heart Disease & Heart Disease & \multicolumn{1}{c}{Total} \\
      \cline{2-5}
      & \makecell{Not Heart \\ Disease} & 657 & 57 & 714 \\
      & \makecell{Heart \\ Disease} & 123 & 633 & 756 \\
      \cline{2-5}
      & Total & 780 & 690 & 1470 \\
    \cline{2-5}
    \end{tabular}
    \subcaption{Validation Data Confusion Matrix}
  \label{tab:validation_confusion_matrix}
\end{table}

\FloatBarrier

\subsection{Hardware Design}
The development of a patient-centric prototype hinged critically on the design of robust hardware. This hardware design imperative necessitated an interface that seamlessly balances user-friendliness while effectively addressing longstanding issues associated with traditional Holter monitors. These challenges encompassed concerns related to skin irritation, timely communication of patient conditions to healthcare providers, sub-optimal accuracy rates, and the inconvenience posed by the cumbersome, outdated hardware apparatus.

\subsubsection{ECG Module}
The AD8232 ECG module was ultimately determined to be the best option for our project's hardware basis after thorough investigation and cautious consideration. This strategic choice was influenced by several compelling elements, all of which added to the module's fit for our project. The AD8232 module stands out for its low power consumption, which is in line with the efficiency goals of our project and provides a practical, affordable option for resource-conscious development. Given the intrinsically weak nature of ECG signals, its outstanding High Common Mode Rejection Ratio (CMRR) is crucial in maintaining the integrity of acquired ECG signals. Additionally, the module's broad input voltage range, high input impedance, and built-in Right-Leg Drive (RLD) circuitry all improve signal quality and patient comfort by successfully reducing problems like skin irritation. Perhaps most importantly, the module can be easily integrated into a variety of wearable health devices, including smartwatches and wristbands, which exactly meet the needs of our project. The AD8232 ECG module is positioned as the cornerstone of our hardware design thanks to this extensive collection of qualities, making it well-suited to carry out the goals of our project.

\subsubsection{Solar-Panel Exterior}
Throughout our research endeavours, our team embarked on a mission to conceptualize and develop an affordable, precise, and high-quality biomedical product that could be accessible to the broader populace, thereby democratizing healthcare access. An intriguing outcome of our research efforts culminated in the exploration of perovskite, a material distinguished by its three-dimensional atomic or ionic arrangement, which may include metal halides. The global appeal of perovskite centres around its exceptional attributes, notably its lightweight nature and cost-effectiveness. Particularly promising are the recent advancements in perovskite-based photovoltaic cells, as exemplified by the remarkable 23.03\% efficiency achieved by researchers from South Korea and Pakistan. This breakthrough underscores the material's viability as a potential replacement for conventional solar cells. Incorporating perovskite into our concept holds the promise of assuring continuous charging, preventing any chance of patient condition degradation during charging intervals, and ensuring uninterrupted gadget performance. This invention perfectly complements our main objective of continuous and seamless patient monitoring.

\subsubsection{ECG Leads}
The integration of adhesive lead patches, widely recognized for their established presence in commercial applications, constitutes a pivotal component of our project implementation. This strategic inclusion serves a dual-fold purpose: first and foremost, it bolsters user convenience, while simultaneously facilitating sustained and effective utilization of the product over the long term. Notably, our project accommodates two distinct variations of these leads to cater to varying patient requirements. One variant seamlessly integrates these leads within our wristband design, while the other introduces an additional adhesive patch on the chest region. This strategic augmentation enhances data acquisition, leading to improved model performance and heightened result accuracy.

\subsubsection{WiFi Data Transfer Module}
Within the scope of our project, the ESP32 module emerged as the optimal choice for our WiFi module, a decision that was substantiated by the outcomes of our extensive research. The ESP32 module was distinguished by several noteworthy attributes that rendered it a standout selection. These attributes encompassed its impressive traits such as low power consumption and cost-effectiveness. Moreover, the module's suitability for the intricate task of ECG data acquisition was underscored by its dual-core processor, affording the capability for simultaneous data capture from our ECG module and seamless data transmission to the cloud. The versatility of the ESP32 module, encompassing both WiFi and Bluetooth connectivity, further bolstered its appeal and utility. Collectively, these attributes established the ESP32 as a premier IoT device, adept at facilitating high-fidelity data transfer and the reliable execution of critical processes.

\subsection{Software}
\subsubsection{Establishment and Maintenance of Cloud Database}
Throughout the execution of our project, the meticulous configuration of resilient infrastructure to seamlessly accommodate patient background data, health records, and real-time ECG data emerged as a paramount consideration. In harmony with the specific prerequisites of our project, we strategically opted to integrate the Amazon Web Services (AWS) IoT Core with AWS for Healthcare and Life Sciences. This deliberate fusion was orchestrated with several pivotal objectives in mind.

Foremost among these objectives was the reinforcement of our project's capabilities in terms of reliable and uninterrupted connectivity. This integration empowered us to achieve frictionless wireless data transfer and secure data storage, effectively leveraging our IoT devices for these purposes. Additionally, the incorporation of AWS for Healthcare and Life Sciences proved instrumental in ensuring adherence to stringent healthcare data privacy regulations, affirming our commitment to maintaining the requisite standards.

Of notable significance was the streamlined implementation of robust encryption protocols and secure storage mechanisms for patient ECG data. These measures, which would otherwise necessitate substantial expertise, time, and computational resources, were seamlessly integrated as a result of this strategic partnership.

Furthermore, this integration furnished our project with the invaluable advantage of real-time data monitoring and visualization, instrumental elements in the successful execution of our objectives.

An additional noteworthy enhancement we consider for our project is the development of a dedicated database, leveraging established frameworks such as MongoDB or MySQL. The adoption of such frameworks holds the potential to significantly enhance our data accessibility and manipulation capabilities to align with our specific project requirements.

\subsubsection{Project Website and App Design}
The development of a user-centric project website and mobile application stands as a pivotal component of our ongoing research initiative. The forthcoming design of the user interface (UI) and user experience (UX) for these digital platforms will be characterized by meticulous attention to detail, with the overarching aim of ensuring effortless interaction and accessibility for both patients and healthcare professionals.

Our vision for the project website is that it will serve as a comprehensive gateway, granting users convenient access to their health records, real-time ECG data, and essential cardiac health information. Similarly, the impending mobile application is designed to empower users with a versatile means to monitor their health while on the move, delivering real-time notifications and intuitive data visualization.

At present, the website is under construction, leveraging a skill set encompassing HTML, CSS, and JavaScript. This strategic choice bestows unparalleled adaptability and affords the website developer full control over its features. Although the development of the mobile application is yet to commence, the plan entails its creation using either Flutter or React Native.

In both cases, data storage and retrieval for the website and app will be facilitated by AWS, a selection motivated by its scalability, robust data security measures, and alignment with healthcare privacy regulations. In essence, our choice of these tools and services was underpinned by their capacity to provide a resilient and secure foundation for our project's digital interface, harmonizing seamlessly with our overarching objectives of accessibility, data protection, and cross-platform usability.

\section{Discussion}
We have attained a commendable accuracy rate of 89\%, leveraging an amalgamation of cost-effective yet high-quality hardware to deliver real-time results of significant merit. Nevertheless, one notable constraint is the restriction on real-time data acquisition from the device. Furthermore, the extended-term utility remains uncertain, as testing duration spanning several weeks or months continue to align with those of a conventional Holter monitor. To address this limitation, we propose a potential enhancement: the training of data on more robust computational infrastructure, possibly incorporating edge machine learning techniques. This augmentation holds promise for further elevating accuracy levels.

\section{Conclusion}
Examination of the results and graphical representations underscores the exceptional performance of our model, particularly in the context of incorporating patient background demographics and medical history. This achievement not only presents a distinctive solution for patients but also offers vital support to healthcare professionals worldwide.

\vspace{12pt}

\end{document}